\documentclass[aps,prl,showpacs,twocolumn,citeautoscript,superscriptaddress,footinbib,bibnotes]{revtex4-1}
\usepackage{graphicx}% Include figure files
\usepackage{dcolumn}% Align table columns on decimal point
\usepackage{bm}% bold math
\usepackage{amsmath}% subequations
\usepackage{color}

\usepackage{hyperref}
\hypersetup{colorlinks=true, linkcolor=blue, citecolor=red, urlcolor=magenta, pdfauthor={Andrea Scaramucci}}

\begin{document}
\preprint{}
\title{Linear Magnetoelectric Effect by Orbital Magnetism}
\author{A. Scaramucci}
\email{andrea.scaramucci@mat.ethz.ch}
\affiliation{Materials Theory, ETH Zurich, Wolfgang-Pauli-Strasse 27, CH-8093 Zurich, Switzerland}
\author{E. Bousquet}
\affiliation{Materials Theory, ETH Zurich, Wolfgang-Pauli-Strasse 27, CH-8093 Zurich, Switzerland}
\affiliation{Physique Th\'{e}orique des Mat\'{e}riaux, Universit\'{e} de Li\`{e}ge, B-4000 Sart Tilman, Belgium}
\author{M. Fechner}
\affiliation{Materials Theory, ETH Zurich, Wolfgang-Pauli-Strasse 27, CH-8093 Zurich, Switzerland}
\author{M. Mostovoy}
\affiliation{Zernike Institute for Advanced Materials, University of Groningen, Nijenborgh 4, 9747AG Groningen, The Netherlands}
\author{N.A. Spaldin}
\affiliation{Materials Theory, ETH Zurich, Wolfgang-Pauli-Strasse 27, CH-8093 Zurich, Switzerland}
%\date{\today}

\begin{abstract}
We use symmetry analysis and first principles calculations to show that the linear magnetoelectric effect can originate from the response of orbital magnetic moments to
the polar distortions induced by an applied electric field. 
Using LiFePO$_4$ as a model compound we show that spin-orbit coupling partially lifts the quenching of the $3$d orbitals and causes small orbital magnetic moments ($\mu_{(L)}\approx 0.3 \mu_B$)
parallel to the spins of the Fe$^{2+}$ ions.
An applied electric field $\mathbf{E}$ modifies the size of these orbital magnetic moments inducing a net magnetization linear in $\mathbf{E}$. 
\end{abstract}


\maketitle
%\section{Introduction}
The last decade has seen increasing interest in the study of coupling between electric polarization and intrinsic magnetic moments
in materials \cite{MFiebig2005}.  
Such {\it magnetoelectric coupling} manifests in numerous macroscopic phenomena: 
Two well known examples are so-called type-II multiferroism \cite{DKhomskii2009} in which the onset of magnetic order induces a spontaneous polarization,
and linear magnetoelectricity, where an applied electric field $\mathbf{E}$ (magnetic field, $\mathbf{H}$), induces a magnetization $M_j=\alpha_{ij} E_i$ (polarization,
$P_i=\alpha_{ij} H_j$).
Although the two phenomena are non-reciprocal, that is many multiferroics do not show a linear magnetoelectric effect and vice-versa, they
are believed to share closely-related microscopic mechanisms.

First-principles computations have been particularly informative in resolving quantitatively the various 
microscopic contributions to magnetoelectric response \cite{JIniguez2008,EBousquet2011,MMostovoy2010}. 
The first study \cite{JIniguez2008} extracted the ``ionic spin'' contribution to $\alpha$, by calculating 
the change in spin canting caused by an $\mathbf{E}$-induced polar distortion
\footnote{Note that, while we use the term ``ionic spin'', this contribution includes couplings such as Dzyaloshinskii-Moriya interaction and electric field dependence of magnetocrystalline anisotropy
both
of which are mediated by spin-orbit interaction, in addition to spin-only contributions from exchange-striction.}.
Subsequently, the methodology to calculate the ``electronic spin'' component was implemented, 
through calculating the electric polarization induced by an applied Zeeman $\mathbf{H}$ field that 
couples only to 
the spin component of the magnetization \cite{EBousquet2011}. In this method, the electronic spin response
is obtained by ``clamping''  the ions during the calculation; relaxing the ionic positions in response
to the $\mathbf{H}$ field yields both the ionic and electronic spin components. 
Interestingly, and perhaps surprisingly, this study showed that the ionic and electronic contributions to 
$\alpha$ can have similar magnitudes. 

\begin{figure}[t]
 \includegraphics[scale=0.085]{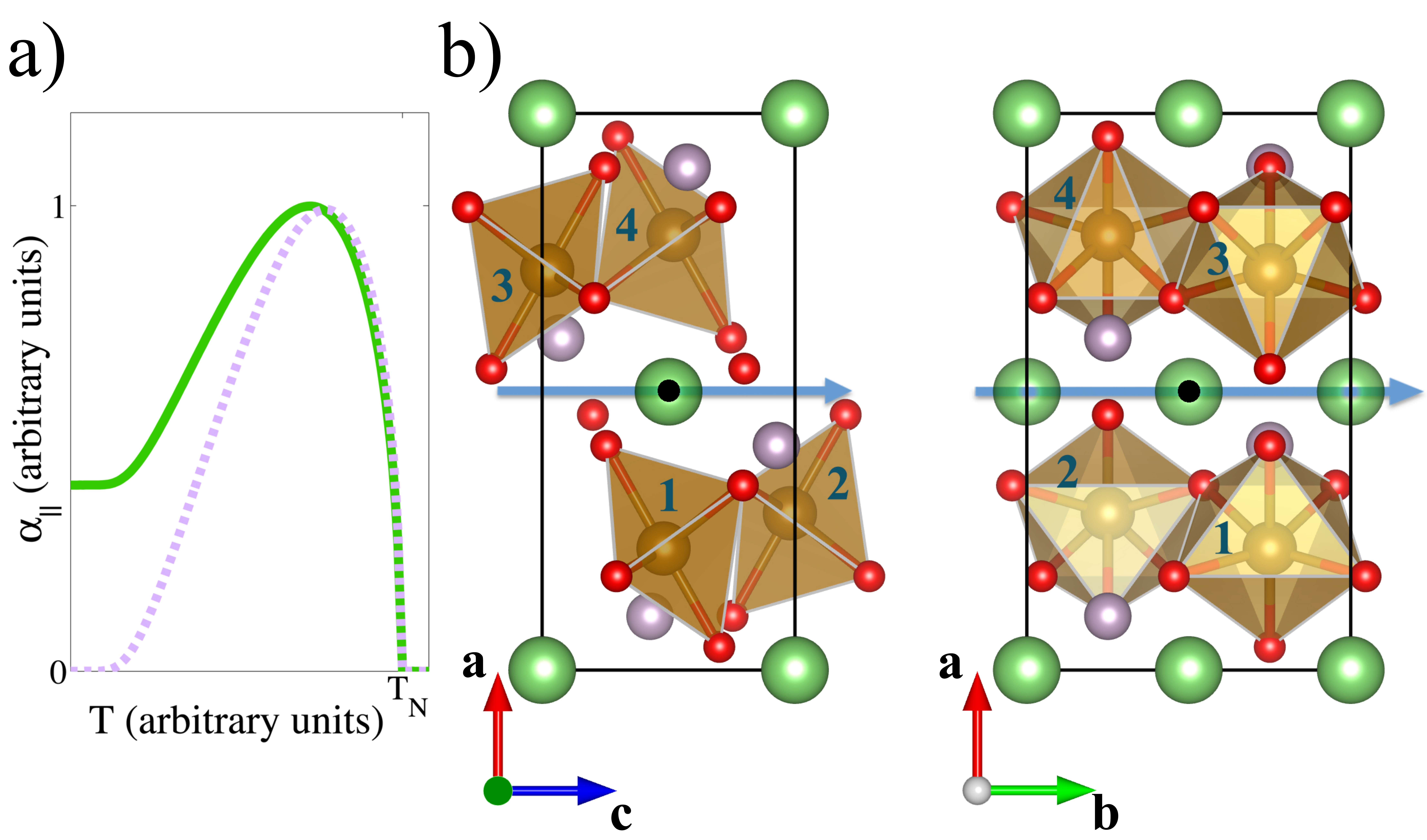}
\caption{a) Qualitative sketches of the temperature dependence of $\alpha_{\parallel}$ in collinear antiferromagnetic magnetoelectrics such as LiFePO$_4$ (solid line) or Cr$_2$O$_3$ (analogous to
this curve but with negative zero temperature value) and that
calculated within a spin-exchange striction mechanism (dashed line).
(b) The orthorhombic unit cell of LiFePO$_4$ contains four Fe$^{2+}$ magnetic cations (brown spheres) which are coordinated by distorted oxygen
(red spheres) octahedra. Li and P ions are represented, respectively, by green and purple spheres. The blue numbers label the magnetic sublattices.
The arrows indicate the screw rotation axis parallel to $\mathbf{b}$ and $\mathbf{c}$ while the black dot indicates the center of inversion.}
 \label{Fig_Structure}
\end{figure}

These spin-based contributions to $\alpha$ have been shown to capture much of the experimental response.
For the case when the magnetic field is applied perpendicular to the spins in a collinear antiferromagnet,
the magnetoelectric coupling, $\alpha_{\perp}$ is relativistic in origin, resulting e.g. from the electric-field dependence of the antisymmetric Dzyaloshinskii-Moriya
exchange \cite{RHornreich1967,MMostovoy2010}. The calculated zero kelvin polarizations are consistent with experimental values \cite{JIniguez2008}, and the temperature evolution of $\alpha_{\perp}$
follows that of the antiferromagnetic order parameter \cite{RHornreich1967}.
The behavior of $\alpha_{\parallel}$ -- obtained when the magnetic field is
applied parallel to the spins -- is more complicated. In this case, the Heisenberg exchange 
interactions between the spins induce an electric polarization at finite temperature which is approximately an
order of magnitude larger than that from the anisotropic exchange interactions of relativistic 
origin responsible for $\alpha_{\perp}$\cite{MMostovoy2010}. 
It has been shown that responses calculated within this
Heisenberg exchange model \cite{MMostovoy2010} agree closely with experiment in the region close to $T_N$  
(Fig.~\ref{Fig_Structure}(a))\footnote{Here, in contrast to Ref.\cite{MMostovoy2010}, we consider mean field theory for a quantum Heisenberg model.}.
One experimentally observed feature is lacking, however: While Heisenberg exchange predicts
$\alpha_{\parallel}\rightarrow 0$ for $T \rightarrow 0$ K, consistent with the vanishing parallel
spin susceptibility at zero kelvin, 
many magnetoelectrics with collinear antiferomagnetism  have non-zero $\alpha_{\parallel}$ at zero kelvin,
and instead follow the temperature dependence sketched in Fig.~\ref{Fig_Structure}(a) (solid line). 
An obvious candidate for the discrepancy is the neglect of orbital contributions \cite{GRado1962}.

While the neglect of orbital magnetism in the above methods is partially justified by the strong quenching of 
$3d$ orbital moments which usually occurs in transition metal oxides, 
spin-orbit coupling,  $H_{so}=\lambda \mathbf{L} \cdot \mathbf{S}$, can of course reduce the quenching, 
and allow a non-negligible orbital magnetization.
This scenario is likely in the collinear antiferromagnet LiFePO$_4$ and in LiCoPO$_4$. 
Both these compounds have a substantially non-zero $\alpha_{\parallel}$ as $T\rightarrow 0$ and an anomalously large anisotropy of the magnetic g-tensor \cite{Creer1970,Liang2008}.

Calculation of the orbital contribution to the magnetoelectric response is not straightforward,
and only a few examples, for limited cases and specific approximations, exist in the literature. 
An early study of LiCoPO$_4$ calculated the ``electronic orbital'' (clamped ion) contribution 
analytically, by determining the change in g-factor with electric field using perturbation theory
within a single-ion Hamiltonian
\cite{IKornev1999}.
While giving a non-zero value for $\alpha_{\parallel}$ at $T=0$, this method underestimated its magnitude.
More recently first-principles 
finite-electric-field methods were used to calculate the electronic orbital contributions to the trace of 
the magnetoelectric tensor -- the  Chern-Simons term --
for Cr$_2$O$_3$ and BiFeO$_3$ \cite{Coh2011,Malashevich2010}. This contribution was shown 
to be negligible with respect to the spin contribution in both cases. 
In this letter we explore the remaining ``ionic orbital'' contribution to the magnetoelectric response by calculating the dependence of the local, on-site orbital magnetic moments on polar lattice
distortions
using density functional theory \cite{Vanderbilt2012}. 
Using magnetoelectric LiFePO$_4$ as a model compound, we show that this ionic orbital contribution to 
$\alpha$ is unexpectedly large and can explain the anomalous low-temperature behavior observed in 
certain components of $\alpha$ that were previously not understood.

LiFePO$_4$ is orthorhombic (space group $Pnma$) and its unit cell (see Fig.~\ref{Fig_Structure}(b)) contains four magnetic sublattices occupied by Fe$^{2+}$ ($S$=2) ions.
Each magnetic ion is surrounded by strongly distorted polar oxygen octahedra for which the only remaining local symmetry is a mirror transformation perpendicular to the
crystallographic axis $\mathbf{b}$ giving local $C_s$ symmetry.
At temperatures below T$_N\approx50$ K the Fe$^{2+}$ magnetic moments order in the antiferromagnetic collinear structure with order parameter
$\mathbf{G}=\mathbf{m}_1-\mathbf{m}_2+\mathbf{m}_3-\mathbf{m}_4$ where $\mathbf{m}_i$ is the magnetization of the $i$-th sublattice. 
The spin orientation in the antiferromagnetic state is still slightly controversial.
Early elastic neutron scattering and X-ray diffraction data suggested that the magnetic moments are fully oriented along the
$\mathbf{b}$ direction \cite{Santoro1967,Streltsov1993}. However, recent neutron scattering measurements
\cite{Li2006} provide evidence for a magnetic structure in which $\mathbf{G}$ is slightly rotated from $\mathbf{b}$.
In this paper, we study only those components allowed with $\mathbf{G}\parallel \mathbf{b}$; $G^a\neq 0$ or $G^c\neq 0$ would give rise to additional non-zero components of the magnetoelectric
tensor \cite{Online} that have not yet been reported.\\
The onset of the antiferromagnetic order breaks inversion symmetry and allows for linear magnetoelectric couplings in the free energy
\begin{equation}
\Phi_{\parallel}= \lambda_{\parallel} G^b  E^a H^b \;\;\;\;  \mbox{and} \;\;\;\;  \Phi_{\perp}=\lambda_{\perp} G^b E^b H^a,
\end{equation}
where $\lambda_i=\alpha_i/G^b$ and the subscript denotes whether the magnetic field is longitudinal or transverse to the collinear magnetic moments.\\

$\alpha_{\parallel}$ follows the typical form discussed previously and sketched in Figure~\ref{Fig_Structure} (a):
Decreasing the temperature from $T_N$, $\alpha_{\parallel}$ rapidly increases and reaches a maximum at $T_{max}\approx45$ K. 
Below $T_{max}$, $\alpha_{\parallel}$  decreases until $20$ K at which it becomes almost temperature independent with 
a value of $\sim 2$ ps$/$m 
\footnote{Since the $\alpha_\parallel(T)$ curve in Ref.~\cite{MMercier1968} is in arbitrary units, we set its maximum to the value of $\alpha^{max}_{\parallel}=10^{-4}$ (Gaussian units)$\approx 4.2$
ps/m from Ref.~\cite{ARomanov2006} in order to estimate $\alpha_\parallel(T=0)$.}. Importantly, it does not approach zero as $T \rightarrow 0$ K.
$\alpha_{\perp}$ has the simpler temperature dependence mentioned earlier, increasing with decreasing temperature below $T_N$ to reach a roughly constant value 
below $25$ K (4 ps$/$m) \cite{MMercier1968}.\\ 

We focus on the microscopic couplings which can induce $\alpha_{\parallel}$.
Phenomenologically, exchange-striction couplings between electric polarization and spins are allowed by symmetry and give rise to the term: $P^a
\propto(\mathbf{m}_1 \cdot \mathbf{m}_3 - \mathbf{m}_2 \cdot \mathbf{m}_4)$ (see Tab.~\ref{tab_PointGroup}). This coupling results in a temperature behavior of
$\alpha_{\parallel}$ similar to that discussed above for Cr$_2$O$_3$ \cite{MMostovoy2010}.
We note that the local symmetry $C_s$ of the crystal field around each Fe$^{2+}$ ion has only one-dimensional irreducible representations and, therefore, the $d$ orbitals are non degenerate.    
When the orbital moments are fully quenched the magnetic moment at the $i$-th site is proportional to the spin
$\mathbf{m}_i=2 \mu_B \mathbf{S}_i$. As discussed above, at $T=0$ the spins in a uniaxial antiferromagnet are not modified by $\mathbf{H}_{\parallel}$ weaker than the magnetic field necessary to flop the spins.
Therefore, the electric polarization generated at $T=0$ by the above couplings in response to $\mathbf{H}_{\parallel}$ is zero.

Next we analyze the orbital contribution to $\alpha_{\parallel}$.
We begin by discussing the orientation and size of orbital moments in zero applied field.
From an atomistic perspective, when $H_{so}=\lambda \mathbf{L} \cdot  \mathbf{S}$ is considered the orbital moments are partially unquenched and the
magnetic moment at site $i$ is:
\begin{equation}
 m^{\mu}_i=\mu_B (2 S^{\mu}_i + L^{\mu}_i)=\mu_B g^{\mu\nu}_i S^{\nu}_i
\end{equation}
where $\mathbf{L}_i$ and $g^{\mu\nu}_i$ are, respectively, the orbital momentum operator and the gyromagnetic tensor at site $i$, $\mu,\nu=a,b,c$ and
summation over repeated indexes is implied. 
For an ion with non-degenerate ground state first-order corrections in $\lambda$ lead to
$g_{\mu\nu}=(2-\lambda \Lambda^{\mu\nu}_i)$ where
$%\begin{equation}
 \Lambda^{\mu\nu}_i=\sum_n \frac{\langle \psi_0 |L^{\mu}| \psi_n \rangle \langle \psi_n |L^{\nu}| \psi_0 \rangle }{\epsilon_n -\epsilon_0}
%\label{Eq_Lambda}
$%\end{equation}
. Here $\psi_0$ is the ground state wave function and $\epsilon_n$ and $\psi_n$ are, respectively, the energy and the wave function of the $n$-th excited state of the Fe$^{2+}$ ion at site i.
Since the magnetic moments are parallel to $\mathbf{b}$ we consider the components  $\Lambda^{\mu b}_i$. 
The transformations of these components under the generators of the space group (modulo primitive translations) are listed in Tab.~\ref{tab_PointGroup}, where we see that
$\Lambda^{ab}_i=\Lambda^{cb}_i=0$ and
$\Lambda^{bb}_i\equiv\Lambda^{bb}$ at every magnetic sublattice \cite{Online}.

\begin{table}[t]
\centering
\begin{tabular}{|c|c|c|c|c|c|c|c|c|c|c|}
\hline
 \,\,\,\,\,\,\,   & \,1\, & \,2\, &  \,3\, & \,4\, & $\Lambda^{ab}_i$ &$\Lambda^{bb}_i$&$\Lambda^{cb}_i$ & \,\,\,\,\,$E^a$ & \,\,\,\,\,$E^b$& \,\,\,\,\,$E^c$ \\
[0.4ex] \hline
  $I$& 4 & 3 & 2 & 1 &  \,\,\,$\Lambda^{ab}_{I(i)}$ &   \,\,\,$\Lambda^{bb}_{I(i)}$&  \,\,\,$\Lambda^{cb}_{I(i)}$ & $-E^a$& $-E^b$& $-E^c$\\
\,\,$2_c$& 2 & 1 & 4 & 3 & \,\,\,\,\,\,$\Lambda^{ab}_{2_z(i)}$ & \,\,\,\,\,\,$\Lambda^{bb}_{2_c(i)}$ &  \,$-\Lambda^{cb}_{2_c(i)}$& $-E^a$& $-E^b$&  \,\,\,\,\,$E^c$\\
\,\,$2_b$& 4 & 3 & 2 & 1 & \,$-\Lambda^{ab}_{2_y(i)}$ &  \,\,\,\,\,\,$\Lambda^{bb}_{2_b(i)}$&  \,$-\Lambda^{cb}_{2_b(i)}$& $-E^a$& $ \,\,\,\,\,E^b$& $-E^c$\\
[0.1ex] \hline
\end{tabular}
\caption{Transformation of the four magnetic sublattices (second to fifth column) under the three generators of the space group (modulo a primitive translation) of LiFePO$_4$: inversion $I$, two fold
screw
rotations around the
$\mathbf{c}$ axis $2_c$, and around the $\mathbf{b}$ axis $2_b$ (see Fig.~\ref{Fig_Structure}). 
Columns six to eight show the transformation of three components of the rank 2 axial tensor $\Lambda_i$ at the $i$-th magnetic sublattice. 
Here the subscripts refer to the change of magnetic sublattice, e.g.  $2_c(\Lambda^{cb}_3)=-\Lambda^{cb}_{2_c(3)}=-\Lambda^{cb}_{4}$. 
The last three columns show the transformations of $\mathbf{E}$.}
\label{tab_PointGroup}
\end{table}

The mean values of the orbital parts of the magnetic moments induced by the antiferromagnetic ordering are $\mu_{(L)i}^\nu= -\lambda \Lambda_i^{\mu b} \langle S_i^b \rangle$. For d$^6$ ions,
$\lambda<0$, therefore
the orbital moment is parallel to the spins in every magnetic sublattice and like the spins, gives rise to zero net magnetic moment.

Next we consider the case $E\neq0$.
Electric-field-induced polar lattice distortions modify the crystal field around each Fe$^{2+}$ ion and the energies
$\epsilon_n=\epsilon_n(\mathbf{E})$. 
Expanding $\Lambda^{\mu\nu}_{i}$ to first order in $\mathbf{E}$ one obtains: $\Lambda_i^{\mu,\nu}(E) = \Lambda_i^{\mu,\nu}(0) +  E^{\alpha} \partial_{E^{\alpha}}\Lambda_i^{\mu,\nu}$, where
\begin{equation}
\partial_{E^{\alpha}}\Lambda^{\mu\nu}_i \!=-\!\!\sum_n\! \frac{\langle \psi_0 |L^{\mu}| \psi_n \rangle \langle
\psi_n |L^{\nu}| \psi_0 \rangle }{(\Delta\epsilon_n)^2} \frac{\partial (\Delta\epsilon_n)}{\partial E^{\alpha} } + \xi^{\mu\nu}_{\rho}\!
\label{Eq_LambdaofE}
\end{equation}
and $\Delta\epsilon_n=\epsilon_n(\mathbf{E})-\epsilon_0(\mathbf{E})$. $ \xi^{\mu\nu}_{\rho}$ are the remaining terms containing derivatives of wave functions with respect to $E^{\alpha}$.
The transformations of the derivatives $\partial_{E^{\alpha}}\Lambda^{\mu b}_i$ under the space group of LiFePO$_4$ can be obtained from those of $\Lambda^{\mu b}_i$ and those of
$\mathbf{E}$ \cite{Online} in Tab.~\ref{tab_PointGroup}.
From these transformations we obtain $\partial_{E^a} \Lambda^{bb}_1=\partial_{E^a} \Lambda^{bb}_3=-\partial_{E^a} \Lambda^{bb}_2=-\partial_{E^a} \Lambda^{bb}_4\equiv\partial_{E^a} \Lambda^{bb}$.
Therefore, the response of the average orbital-induced magnetic moment to an electric field along $\mathbf{a}$ gives rise to a net magnetization along $\mathbf{b}$ 
\begin{equation}
\mu^b_{(L)} = \mu_B \partial_{E^a} \Lambda^{bb} (\langle S^b_1\rangle-\langle S^b_2\rangle+\langle S^b_3\rangle -\langle S^b_4 \rangle) E^a
\label{Eq_OrbMagResp}
\end{equation}
that at $T=0$ gives $\mu^b_{(L)}  = 4  \mu_B S \partial_{E^a} \Lambda^{bb} E^a$.\\

To calculate the strength of the linear magnetoelectric coupling arising from this mechanism, we perform first principles calculations using the Vienna {\it ab initio} simulation package (VASP) \cite{Kresse1}. We use a plane-wave
basis set for
the expansion of the electronic valence wave function and PAW \cite{Kresse2} potentials for the treatment of core electrons. The
exchange-correlation potential is described within the local-spin-density approximation plus a rotationally invariant Hubbard-$U$ (LSDA+$U$)
with a $U$ value of 5 eV, and $J$ values between 0 and 1 eV.
Calculations are performed at the experimental unit cell volume of $291$ \AA{}$^3$ \cite{Streltsov1993}.
We first relax the structure in the absence of spin-orbit coupling and then we include spin-orbit coupling to calculate the orbital magnetic moment. We obtain an orbital moment  
$\mu_{(L)} = 0.306 \mu_B$ parallel to the spins when we use a $J$ value of 1eV. We note that the magnitude of the magnetic moment depends on $J$ and on the PAW sphere radius used as discussed in
Ref.~\cite{Online}.\\

To calculate the ionic orbital response -- that is the change in orbital magnetic moments when the ions are displaced by an applied electric field -- we adapt the framework introduced in Ref.~\cite{JIniguez2008} to obtain the ionic spin response. 
As in Ref.~\cite{JIniguez2008}, we shift 
the equilibrium positions $\mathbf{r}_i$ of the ions by $\Delta r^{\mu}_i = E^{\rho} \sum_{\nu j} \phi^{-1}_{\mu i, \nu j} Z^{\ast}_{j,\rho
\nu} $ where 
 $\phi^{-1}_{\mu i, \nu j}$ is the inverse of the force constant matrix after the acoustic modes are traced out and $Z^{\ast}_{j,\rho \nu}$ are the Born effective charges, both calculated in the absence of spin-orbit coupling.
\begin{figure}[t]
 \includegraphics[scale=0.4]{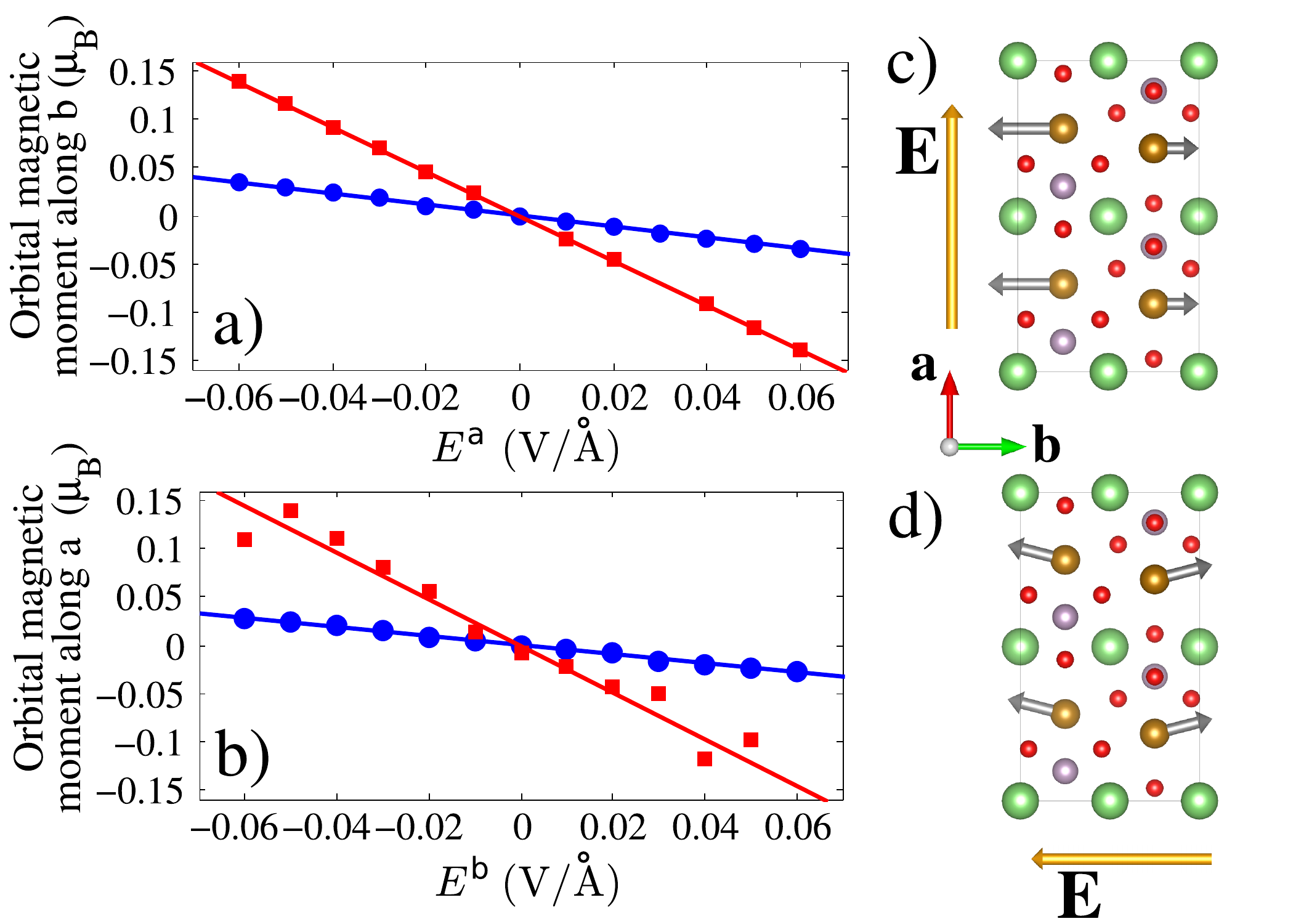}
\caption{Calculated electric-field dependence of the net orbital magnetic moment per unit cell. $\mathbf{E} \parallel \mathbf{a}$ (panel a)) results in an
orbital magnetization along $\mathbf{b}$ ($\alpha_{\parallel}$) while $\mathbf{E}\parallel \mathbf{b}$ (panel
b)) produces a net orbital magnetic moment along $\mathbf{a}$ ($\alpha_{\perp}$). Blue dots and red squares are calculated values
at J=1 eV and J=0 eV respectively, while the straight lines are linear fits to the calculated values. 
The cartoons on the right panels show the size and orientation of the orbital magnetic moments (gray arrows) of Fe$^{2+}$ (brown spheres) when the electric field (yellow arrow) is applied
along $\mathbf{a}$ (panel c) and $\mathbf{b}$ (panel d). In the cartoon the size of the effect is increased for clarity.}
 \label{Fig_MEeffect}
\end{figure}
Since we aim to separate the orbital from the spin contribution, we constrain the orientation of the spins to lie along the 
$\mathbf{b}$ direction, which we call the ``clamped spin'' approximation (note, however that the magnitude of the spin is
unconstrained.) After making the $\Delta r^{\mu}_i$ distortions from the equilibrium zero-field positions, we relax the
electronic density with spin-orbit coupling included and calculate the resulting orbital magnetic moments. 

Figure \ref{Fig_MEeffect}(a) shows the evolution of the calculated net orbital magnetic moment $\mu_{(L)}=\sum_{i=1,4} \mu_{(L),i}$ of one unit cell of LiFePO$_4$ for an electric field applied along
$\mathbf{a}$ with $J=1$ eV (blue points) and $J=0$ eV (red points). 
(Note that, while the electric field is applied perpendicular to the spins, this corresponds to
the parallel component of $\alpha$, since the magnetoelectric response is off-diagonal). 
We find that at non-zero electric field the orbital magnetic moments remain parallel to the spins, and consistent with Eq.~(\ref{Eq_OrbMagResp})
the change of their size is opposite for odd and even magnetic sublattices giving rise to a net magnetization.
The linear fits of the $E^a$ responses of the orbital magnetization at $J=1$ eV (blue line) and $J=0$ eV (red line) give $\alpha_{\parallel}=2.3$ ps$/$m and  $\alpha_{\parallel}=9.3$
ps$/$m respectively.
The $\alpha_{\parallel}$ value for $J=1$ eV is reasonably close to the experimental value of $\alpha_{\parallel} \sim 2$ ps/m at $T=0$ K \cite{MMercier1968,ARomanov2006}. This value of $J$ is
consistent with
Ref.~\cite{EBousquet2010} which showed that it is necessary to use 
$J>0.6$ eV to obtain the correct magnetic easy axis. To summarize this section, we find that the 
calculated zero kelvin ionic orbital contribution to $\alpha_{\parallel}$ has a value which is 
consistent with the measured value of $\alpha_{\parallel}$. We suggest, therefore, that the 
previous discrepancy between the measured zero kelvin magnetoelectric response and the calculated
spin-only response can be explained by this contribution. 
At non-zero temperatures, contributions to $\alpha_{\parallel}$ that are inactive in the absence of thermal fluctuations, have to be taken into account. 
These terms comprise the electric field dependence of single-ion anisotropy, which has the same nature as the orbital magnetic moment, as well as the Heisenberg interactions mentioned earlier.

Finally, we investigate the ionic orbital contribution to $\alpha_{\perp}$, by calculating 
the effect of an electric field applied along $\mathbf{b}$. While the spin-only contribution
was not inconsistent with experiment in this case, 
contributions to $\alpha_{\perp}$ from the electric field dependence of 
$\mathbf{\mu}_{(L)i}$  have not been previously investigated and might also play a role.
First we use similar symmetry arguments as those used for $\alpha_{\parallel}$ to find constraints on $\partial_{E^b}\Lambda^{\mu\nu}_i$. 
From Tab.~\ref{tab_PointGroup} we find: 
 $\partial_{E^b} \Lambda^{ab}_1=\partial_{E^b} \Lambda^{ab}_3=-\partial_{E^b} \Lambda^{ab}_2=-\partial_{E^b} \Lambda^{ab}_4\equiv\partial_{E^b} \Lambda^{ab}$,
 $\partial_{E^b} \Lambda^{cb}_1=\partial_{E^b} \Lambda^{cb}_2=-\partial_{E^b} \Lambda^{cb}_3=-\partial_{E^b} \Lambda^{cb}_4\equiv\partial_{E^b} \Lambda^{cb}$ and
 $\partial_{E^b} \Lambda^{bb}_i=0$.
On one hand, we note that the transformation properties of $\partial_{E^b} \Lambda^{ab}_i$ are identical to those of $\partial_{E^a} \Lambda^{bb}_i$. This allows
for a linear dependence of the orbital magnetization along $\mathbf{a}$ when the electric field is applied along
$\mathbf{b}$: $\mu_a=4\mu_B E^b \partial_{E^b} \Lambda^{ab} |\langle S^b \rangle|$ where $| \langle S^b \rangle|$ is the absolute value of the average spin component along $\mathbf{b}$.
In contrast, the transformation properties of $\partial_{E^b} \Lambda^{cb}_i$, together with the spin ordering of LiFePO$_4$ show that the change in orbital moment along $\mathbf{c}$ under an
applied $E^b$ field have opposite sign for sublattices $1$, $4$ compared with $2$,$3$, yielding zero net moment in this direction.\\
To obtain the size of the ionic orbital contribution to $\alpha_{\perp}$ we perform {\it ab initio} calculations using the same method discussed for $\alpha_{\parallel}$ but with the electric field applied along
$\mathbf{b}$.
As before, we adopt the clamped-spin approach, and constrain the spins in the $\mathbf{b}$ direction. 
The resulting calculated values of net orbital magnetic moment are shown in Fig.~\ref{Fig_MEeffect}(b) as a function of $E^b$. Here blue and red points show results for, respectively, $J=1$ eV and
$J=0$ eV.
Even when the spins are constrained to be parallel to the $\mathbf{b}$ axes, the applied $\mathbf{E}_{\parallel}$ induces a canting of the orbital magnetic moments from the $b$ direction.
In agreement with the  constraints found for  $\partial_{E^b} \Lambda^{ab}_i$ the resulting canting is uniform along the $\mathbf{a}$ axis for all magnetic sublattices giving rise to a net
magnetization linear in $E_b$.
Furthermore, as predicted using the transformations of $\partial_{E^b} \Lambda^{cb}_i$ for finite $E_b$ we observe a tiny staggered canting  of the orbital moment along $c$ which gives rise to zero
net
magnetization.
The solid lines in Fig.~\ref{Fig_MEeffect}(b) are linear interpolations of the calculated values and give linear magnetoelectric responses of $1.9$ ps$/$m and $9.7$ ps$/$m for $J=1$ eV and $J=0$
respectively.
To these values, which contain only the ionic orbital magnetoelectric effect, one should add the spin-only contribution to $\alpha_\perp$, which in contrast to the case of $\alpha_\parallel$ does not
vanish at $T = 0$.
These include the rotation of easy axis anisotropy, that shares the same origin as the canting of orbital magnetic moment, as well as Dzyaloshinskii-Moriya interaction.
Using the approach described in Ref.~\cite{EBousquet2011}, which includes these contributions but not the orbital moment part, we obtain for $J=1$ eV a value for $\alpha_{\perp}$ of $2.6$ ps$/$m with
sign opposite to the orbital one.
Importantly, these considerations can also be used to describe resonant excitation of waves of oscillating magnetization $M \parallel a$ with an oscillating electric field of a light wave 
$E \parallel b$, resulting in the so-called "electromagnon" peaks in optical absorption \cite{Pimenov2006}. Thus the coupling between the orbital
magnetic moment and electric field gives rise to both static and dynamic magnetoelectric effects. 

In summary, we have shown that a linear magnetoelectric effect can arise from the dependence of orbital magnetic moments on the polar distortions induced by an applied electric field, the so-called ``ionic orbital'' contribution to the magnetoelectric response.
We presented a symmetry analysis which allows the components of $\alpha_{\mu\nu}$ for which this effect exists to be determined, and a methodology which can be used to calculate {\it ab initio} those
components at $T=0$.
We applied the methodology to LiFePO$_4$ and resolved the previous discrepancy between previous calculations of the spin-only contributions
and experiment for $\alpha_{\parallel}$.
Our results show that the orbital contributions to the magnetoelectric response can be comparable in size to the spin contributions
of either relativistic or exchange-striction origin in $3d$
transition metal compounds.
As suggested by Eq.~(\ref{Eq_OrbMagResp}), the temperature dependence of the magnetoelectric effect caused by orbital magnetism coincides with that of the order parameter which,
added to the temperature dependence of magnetoelectric effect originated by striction gives rise to a qualitative agreement for various collinear antiferromagnets such as Cr$_2$O$_3$ \cite{EKita1979},
LiCoPO$_4$ \cite{JRivera1994}  and TbPO$_4$ \cite{GRado1984}.\\
Furthermore, we note that if such coupling between orbital magnetization and polar distortion is allowed by symmetry, its strength does not depend solely on the strength of the spin-orbit interaction.
As shown in Eq.~(\ref{Eq_LambdaofE}), from a single ion perspective, the strength of such an effect is determined by the energy gap between the ground state and the excited states for which $\langle
\psi_0
|L^{\mu}| \psi_n \rangle\neq 0$ and also by the dependence of the energies of ionic orbitals on polar distortions of the crystal field. 
This suggests that large magnetoelectric effect due to the orbital moment correlates with the enhanced anisotropic g-tensor and the anisotropy of the magnetic susceptibility in the paramagnetic state.
In particular, large response of orbital magnetism to an applied electric field might be found in compounds with reasonably small electronic gap, containing magnetic ions with large spin-orbit
coupling and with low symmetry polar oxygen coordination.

This work was supported by ETH Z\"{u}rich and by the European Research Council Advanced Grants program under the FP7, grant number 291151. E. B. thanks FRS-FNRS Belgium for support. 

\bibliography{Biblio}{}

\end{document}

\preprint{}
\title{Supplementary Material for ``Linear Magnetoelectric Effect by Orbital Magnetism''}
\author{A. Scaramucci}
\affiliation{Materials Theory, ETH Zurich, Wolfgang-Pauli-Strasse 27, CH-8093 Zurich, Switzerland}
\author{E. Bousquet}
\affiliation{Materials Theory, ETH Zurich, Wolfgang-Pauli-Strasse 27, CH-8093 Zurich, Switzerland}
\affiliation{Physique Th\'{e}orique des Mat\'{e}riaux, Universit\'{e} de Li\`{e}ge, B-4000 Sart Tilman, Belgium}
\author{M. Fechner}
\affiliation{Materials Theory, ETH Zurich, Wolfgang-Pauli-Strasse 27, CH-8093 Zurich, Switzerland}
\author{M. Mostovoy}
\affiliation{Zernike Institute for Advanced Materials, University of Groningen, Nijenborgh 4, 9747AG Groningen, The Netherlands}
\author{N.A. Spaldin}
\affiliation{Materials Theory, ETH Zurich, Wolfgang-Pauli-Strasse 27, CH-8093 Zurich, Switzerland}
%\date{\today}

\maketitle

{\it Effect of small rotations of $\mathbf{G}$ away from $\mathbf{b}$.} 
The transformations of the magnetic field ($\mathbf{H}$), the order parameter ($\mathbf{G}$) and the electric field
($\mathbf{E}$) under the generators of $Pnma$ (modulo primitive translations) are
shown
in Tab.~\ref{tab_HL} and in Tab.~I in the text.
The invariant terms in the free energy, $\Phi_{ME}$, leading to linear magnetoelectric couplings can be written as
\begin{eqnarray}
\Phi_{ME} 
&=&  G^a (\lambda^a_{\parallel} H^a E^a + \lambda^a_{\perp 1} H^b E^b + \lambda^a_{\perp 2} H^c E^c )  \nonumber \\
& & + G^b (\lambda^b_{\parallel} H^b E^a + \lambda^b_{\perp } H^a E^b ) \nonumber \\ 
& & + G^c (\lambda^b_{\parallel} H^c E^a + \lambda^c_{\perp } H^a E^c ),
\label{Eq_genPhi}
\end{eqnarray}
where $\lambda^i_{\parallel}$ and $\lambda^i_{\perp}$ are functions of temperature.
From Eq.~(\ref{Eq_genPhi}) it is clear that different components of the magnetoelectric tensor correspond to different components of $\mathbf{L}$ along the crystallographic axes.
Therefore, a small rotation of $\mathbf{G}$ from the $\mathbf{b}$ direction does not affect the analysis of $\alpha_{\parallel}$ and $\alpha_{\perp}$ performed in the text
with the assumption of $\mathbf{G} \parallel \mathbf{b}$.\\
\begin{table}[bp]
\centering
\begin{tabular}{|c|c|c|c|}
\hline
  & $I$ & $2_z$ &  $2_y$\\
[0.4ex] \hline
\,\,\, $H^a$\,\,\,&\,\,\, $H^a $\,\,\, & \,\,\,$-H^a$ \,\,\,& \,\,\,$-H^a$\,\,\,\\
\,\,\, $H^b$\,\,\,&\,\,\, $H^b $\,\,\, & \,\,\,$-H^b$ \,\,\,& \,\,\,$H^b $\,\,\,\\
\,\,\, $H^c$\,\,\,&\,\,\, $H^c $\,\,\, & \,\,\,$H^c$ \,\,\,& \,\,\,$-H^c $\,\,\,\\
[0.1ex] \hline
\hline
\,\,\, $L^a$\,\,\,&\,\,\, $-L^a $\,\,\, & \,\,\,$L^a$ \,\,\,& \,\,\,$L^a$\,\,\,\\
\,\,\, $L^b$\,\,\,&\,\,\, $-L^b $\,\,\, & \,\,\,$L^b$ \,\,\,& \,\,\,$-L^b $\,\,\,\\
\,\,\, $L^c$\,\,\,&\,\,\, $-L^c $\,\,\, & \,\,\,$-L^c$ \,\,\,& \,\,\,$L^c $\,\,\,\\
\hline
\end{tabular}
\caption{Transformation of the magnetic field $\mathbf{E}$ and the order parameter under the generators of the point group of $Pnma$.}
\label{tab_HL}
\end{table}
{\it Symmetry analysis of $\Lambda^{\alpha b}_i$ and $\partial_{E^{\mu}}\Lambda^{\alpha b}_i$ extended to three electric field orientations.}
The effect of the generators of $Pnma$  on $\Lambda^{\mu\nu}_i$
can be obtained considering the transformations of the magnetic sublattices (see Tab.~I in the text) and the transformations of
$\mathbf{L}$.
Table~\ref{tab_Lambda} shows the way in which $\Lambda^{\mu\nu}_i$ changes under these generators (modulo primitive translations).
These transformations imply constraints on some of the $\Lambda^{\mu,\nu}_i$ components at each magnetic sublattice. 
For instance, the invariance under inversion, $I$, gives: $\Lambda^{a b}_1 = \Lambda^{a b}_4$ and the invariance under two-fold rotation around $\mathbf{b}$ implies: $\Lambda^{a b}_1 =- \Lambda^{a
b}_4$ which can be both satisfied only for $\Lambda^{a b}_1 = \Lambda^{a b}_4 = 0$.
Similarly, using Tab.~\ref{tab_Lambda}, one finds: $\Lambda^{a b}_2 = \Lambda^{a b}_3 = 0$, $\Lambda^{b b}_1 = \Lambda^{b b}_2=\Lambda^{b b}_3 = \Lambda^{b b}_4$ and 
$\Lambda^{c b}_1 = \Lambda^{c b}_2=\Lambda^{c b}_3 = \Lambda^{c b}_4=0$.\\
A small applied electric field affects the orbital moments by changing $\Lambda^{\mu \nu}_i$. 
The expansion of orbital magnetic moment at the $i$-th sublattice, $\mu^{\alpha}_{(L)i}$, to the linear order in $\mathbf{E}$ gives
\begin{equation}
\mu^{\alpha}_{(L)i}\!=\!\!- \lambda \mu_B \!\!\left(\! \Lambda_i^{\alpha b}(0)+\!\! \sum_{\mu=a,b,c}\!\! \frac{\partial \Lambda_i^{\alpha b}}{\partial E^{\mu}}  E^{\mu} +\! ... \!\right)\!\!
\langle S^b \rangle ,
\label{Eq_OL_MuL}
\end{equation}
where the second term inside the brackets contains the derivatives given in Eq.~(3) in the text.\\
The relative changes of orbital magnetic moments on different sublattice due to an applied electric field are determined by the symmetry properties of $\frac{\partial \Lambda_i^{\alpha
b}} {\partial E^{\mu}}$.
These are listed in Tab.~\ref{tab_Lambda} and are obtained by combining transformations of $\mathbf{E}$ with those of $\Lambda^{\mu \nu}_i$ under the generators of $Pnma$.
Similarly to the case of $\Lambda^{\mu b}_i$, symmetries impose constraints on $\frac{\partial \Lambda^{\alpha b}_i} {\partial E^{\mu}}$.
For example, the set of equations: $\partial_{E^a} \Lambda^{ab}_1=-\partial_{E^a} \Lambda^{ab}_4$ and
 $\partial_{E^a} \Lambda^{ab}_1=\partial_{E^a} \Lambda^{ab}_4$, which need to be satisfied to ensure, respectively, the invariance under inversion and two fold
rotation along $\mathbf{b}$ (see Tab.~\ref{tab_Lambda}), imply that $\partial_{E^a} \Lambda^{ab}_1=\partial_{E^a} \Lambda^{ab}_4=0$.
Analogously, we obtain that $\partial_{E^a} \Lambda^{ab}$, $\partial_{E^a} \Lambda^{cb}$,
$\partial_{E^b} \Lambda^{bb}$, $\partial_{E^c} \Lambda^{ab}$ and $\partial_{E^a} \Lambda^{ab}$ have to vanish at each magnetic sublattice.
Furthermore, the non-vanishing terms must satisfy:
\begin{eqnarray}
 \partial_{E^a} \Lambda^{bb}_1 &=& - \partial_{E^a} \Lambda^{bb}_2 = \partial_{E^a} \Lambda^{bb}_3 =  -\partial_{E^a} \Lambda^{bb}_4  \nonumber \\
 \partial_{E^b} \Lambda^{ab}_1 &=& - \partial_{E^b} \Lambda^{ab}_2 = \partial_{E^b} \Lambda^{ab}_3 =  -\partial_{E^b} \Lambda^{ab}_4  \nonumber \\
 \partial_{E^b} \Lambda^{cb}_1 &=& \partial_{E^b} \Lambda^{cb}_2 = - \partial_{E^b} \Lambda^{cb}_3 =  -\partial_{E^b} \Lambda^{cb}_4  \nonumber \\
 \partial_{E^c} \Lambda^{bb}_1 &=& \partial_{E^c} \Lambda^{bb}_2 = - \partial_{E^c} \Lambda^{bb}_3 =  -\partial_{E^c} \Lambda^{bb}_4 . 
\label{Eq_OL_DLambda} 
\end{eqnarray}
\begin{table}[hp]
\centering
\begin{tabular}{|c|c|c|c|}
\hline
  & $I$ & $2_c$ &  $2_b$\\
[0.4ex] \hline
\,\,\, $\Lambda_1^{ab} $\,\,\,&\,\,\, $\Lambda_4^{ab} $\,\,\, & \,\,\,$\Lambda_2^{ab} $ \,\,\,& \,\,\,$-\Lambda_4^{ab} $\,\,\,\\
$\Lambda_1^{bb} $& $\Lambda_4^{bb} $ & $\Lambda_2^{bb} $ & $\Lambda_4^{bb} $\\
$\Lambda_1^{cb} $& $\Lambda_4^{cb} $ & $-\Lambda_2^{cb} $ & $-\Lambda_4^{cb} $\\
[0.1ex] \hline
\,\,\, $\Lambda_2^{ab} $\,\,\,&\,\,\, $\Lambda_3^{ab} $\,\,\, & \,\,\,$\Lambda_1^{ab} $ \,\,\,& \,\,\,$-\Lambda_3^{ab} $\,\,\,\\
$\Lambda_2^{bb} $& 		      $\Lambda_3^{bb} $ & 	      $\Lambda_1^{bb} $ & 	      $\Lambda_3^{bb} $\\
$\Lambda_2^{cb} $& 		      $\Lambda_3^{cb} $ & 	      $-\Lambda_1^{cb} $ & 	      $-\Lambda_3^{cb} $\\
[0.1ex] \hline
\,\,\, $\Lambda_3^{ab} $\,\,\,&\,\,\, $\Lambda_2^{ab} $\,\,\, & \,\,\,$\Lambda_4^{ab} $ \,\,\,& \,\,\,$-\Lambda_2^{ab} $\,\,\,\\
$\Lambda_3^{bb} $& 		      $\Lambda_2^{bb} $ & 	      $\Lambda_4^{bb} $ & 	      $\Lambda_2^{bb} $\\
$\Lambda_3^{cb} $& 		      $\Lambda_2^{cb} $ & 	      $-\Lambda_4^{cb} $ & 	      $-\Lambda_2^{cb} $\\
[0.1ex] \hline
\,\,\, $\Lambda_4^{ab} $\,\,\,&\,\,\, $\Lambda_1^{ab} $\,\,\, & \,\,\,$\Lambda_3^{ab} $ \,\,\,& \,\,\,$-\Lambda_1^{ab} $\,\,\,\\
$\Lambda_4^{bb} $& 		      $\Lambda_1^{bb} $ & 	      $\Lambda_3^{bb} $ & 	      $\Lambda_1^{bb} $\\
$\Lambda_4^{cb} $& 		      $\Lambda_1^{cb} $ & 	      $-\Lambda_3^{cb} $ & 	      $-\Lambda_1^{cb} $\\
[0.1ex] 
\hline
\hline
\,\,\, $\partial_{E^a} \Lambda_1^{ab} $\,\,\,&\,\,\, $-\partial_{E^a} \Lambda_4^{ab} $\,\,\, & \,\,\,$-\partial_{E^a} \Lambda_2^{ab} $ \,\,\,&\,\,\,$\partial_{E^a} \Lambda_4^{ab} $\,\,\,\\
       $\partial_{E^a} \Lambda_1^{bb} $&             $-\partial_{E^a} \Lambda_4^{bb} $ & 	    $-\partial_{E^a} \Lambda_2^{bb} $ & 	  $-\partial_{E^a} \Lambda_4^{bb} $\\
       $\partial_{E^a} \Lambda_1^{cb} $& 	     $-\partial_{E^a} \Lambda_4^{cb} $ &	     $\partial_{E^a} \Lambda_2^{cb} $ & 	   $\partial_{E^a} \Lambda_4^{cb} $\\
[0.1ex] \hline
$\partial_{E^a}\Lambda_2^{ab} $&                      $-\partial_{E^a}\Lambda_3^{ab} $ &               $-\partial_{E^a}\Lambda_1^{ab} $ &    $\partial_{E^a}\Lambda_3^{ab} $\\
$\partial_{E^a}\Lambda_2^{bb} $& 		      $-\partial_{E^a}\Lambda_3^{bb} $ & 	      $-\partial_{E^a}\Lambda_1^{bb} $ &   $-\partial_{E^a}\Lambda_3^{bb} $\\
$\partial_{E^a}\Lambda_2^{cb} $& 		      $-\partial_{E^a}\Lambda_3^{cb} $ & 	      $\partial_{E^a}\Lambda_1^{cb} $ &	   $\partial_{E^a}\Lambda_3^{cb} $\\
[0.1ex] \hline
$\partial_{E^a}\Lambda_3^{ab} $&                      $-\partial_{E^a}\Lambda_2^{ab} $ &               $-\partial_{E^a}\Lambda_4^{ab} $ &    $\partial_{E^a}\Lambda_2^{ab} $\\
$\partial_{E^a}\Lambda_3^{bb} $& 		      $-\partial_{E^a}\Lambda_2^{bb} $ & 	      $-\partial_{E^a}\Lambda_4^{bb} $ &   $-\partial_{E^a}\Lambda_2^{bb} $\\
$\partial_{E^a}\Lambda_3^{cb} $& 		      $-\partial_{E^a}\Lambda_2^{cb} $ & 	      $\partial_{E^a}\Lambda_4^{cb} $ &	   $\partial_{E^a}\Lambda_2^{cb} $\\
[0.1ex] \hline
$\partial_{E^a}\Lambda_4^{ab} $&                      $-\partial_{E^a}\Lambda_1^{ab} $ &               $-\partial_{E^a}\Lambda_3^{ab} $ &    $\partial_{E^a}\Lambda_1^{ab} $\\
$\partial_{E^a}\Lambda_4^{bb} $& 		      $-\partial_{E^a}\Lambda_1^{bb} $ & 	      $-\partial_{E^a}\Lambda_3^{bb} $ &   $-\partial_{E^a}\Lambda_1^{bb} $\\
$\partial_{E^a}\Lambda_4^{cb} $& 		      $-\partial_{E^a}\Lambda_1^{cb} $ & 	      $\partial_{E^a}\Lambda_3^{cb} $ &	   $\partial_{E^a}\Lambda_1^{cb} $\\
[0.1ex]
\hline
\hline
\,\,\, $\partial_{E^b} \Lambda_1^{ab} $\,\,\,&\,\,\, $-\partial_{E^b} \Lambda_4^{ab} $\,\,\, & \,\,\,$-\partial_{E^b} \Lambda_2^{ab} $ \,\,\,&\,\,\,$-\partial_{E^b} \Lambda_4^{ab} $\,\,\,\\
       $\partial_{E^b} \Lambda_1^{bb} $&             $-\partial_{E^b} \Lambda_4^{bb} $ & 	    $-\partial_{E^b} \Lambda_2^{bb} $ & 	  $\partial_{E^b} \Lambda_4^{bb} $\\
       $\partial_{E^b} \Lambda_1^{cb} $& 	     $-\partial_{E^b} \Lambda_4^{cb} $ &	     $\partial_{E^b} \Lambda_2^{cb} $ & 	   $-\partial_{E^b} \Lambda_4^{cb} $\\
[0.1ex] \hline
$\partial_{E^b}\Lambda_2^{ab} $&                      $-\partial_{E^b}\Lambda_3^{ab} $ &               $-\partial_{E^b}\Lambda_1^{ab} $ &    $-\partial_{E^b}\Lambda_3^{ab} $\\
$\partial_{E^b}\Lambda_2^{bb} $& 		      $-\partial_{E^b}\Lambda_3^{bb} $ & 	      $-\partial_{E^b}\Lambda_1^{bb} $ &   $\partial_{E^b}\Lambda_3^{bb} $\\
$\partial_{E^b}\Lambda_2^{cb} $& 		      $-\partial_{E^b}\Lambda_3^{cb} $ & 	      $\partial_{E^b}\Lambda_1^{cb} $ &	   $-\partial_{E^b}\Lambda_3^{cb} $\\
[0.1ex] \hline
$\partial_{E^b}\Lambda_3^{ab} $&                      $-\partial_{E^b}\Lambda_2^{ab} $ &               $-\partial_{E^b}\Lambda_4^{ab} $ &    $-\partial_{E^b}\Lambda_2^{ab} $\\
$\partial_{E^b}\Lambda_3^{bb} $& 		      $-\partial_{E^b}\Lambda_2^{bb} $ & 	      $-\partial_{E^b}\Lambda_4^{bb} $ &   $\partial_{E^b}\Lambda_2^{bb} $\\
$\partial_{E^b}\Lambda_3^{cb} $& 		      $-\partial_{E^b}\Lambda_2^{cb} $ & 	      $\partial_{E^b}\Lambda_4^{cb} $ &	   $-\partial_{E^b}\Lambda_2^{cb} $\\
[0.1ex] \hline
$\partial_{E^b}\Lambda_4^{ab} $&                      $-\partial_{E^b}\Lambda_1^{ab} $ &               $-\partial_{E^b}\Lambda_3^{ab} $ &    $-\partial_{E^b}\Lambda_1^{ab} $\\
$\partial_{E^b}\Lambda_4^{bb} $& 		      $-\partial_{E^b}\Lambda_1^{bb} $ & 	      $-\partial_{E^b}\Lambda_3^{bb} $ &   $\partial_{E^b}\Lambda_1^{bb} $\\
$\partial_{E^b}\Lambda_4^{cb} $& 		      $-\partial_{E^b}\Lambda_1^{cb} $ & 	      $\partial_{E^b}\Lambda_3^{cb} $ &	   $-\partial_{E^b}\Lambda_1^{cb} $\\
[0.1ex] 
\hline
\hline
\,\,\, $\partial_{E^c} \Lambda_1^{ab} $\,\,\,&\,\,\, $-\partial_{E^c} \Lambda_4^{ab} $\,\,\, & \,\,\,$\partial_{E^c} \Lambda_2^{ab} $ \,\,\,&\,\,\,$\partial_{E^c} \Lambda_4^{ab} $\,\,\,\\
       $\partial_{E^c} \Lambda_1^{bb} $&             $-\partial_{E^c} \Lambda_4^{bb} $ & 	    $\partial_{E^c} \Lambda_2^{bb} $ & 	  $-\partial_{E^c} \Lambda_4^{bb} $\\
       $\partial_{E^c} \Lambda_1^{cb} $& 	     $-\partial_{E^c} \Lambda_4^{cb} $ &	     $-\partial_{E^c} \Lambda_2^{cb} $ & 	   $\partial_{E^c} \Lambda_4^{cb} $\\
[0.1ex] \hline
$\partial_{E^c}\Lambda_2^{ab} $&                      $-\partial_{E^c}\Lambda_3^{ab} $ &               $\partial_{E^c}\Lambda_1^{ab} $ &    $\partial_{E^c}\Lambda_3^{ab} $\\
$\partial_{E^c}\Lambda_2^{bb} $& 		      $-\partial_{E^c}\Lambda_3^{bb} $ & 	      $\partial_{E^c}\Lambda_1^{bb} $ &   $-\partial_{E^c}\Lambda_3^{bb} $\\
$\partial_{E^c}\Lambda_2^{cb} $& 		      $-\partial_{E^c}\Lambda_3^{cb} $ & 	      $-\partial_{E^c}\Lambda_1^{cb} $ &	   $\partial_{E^c}\Lambda_3^{cb} $\\
[0.1ex] \hline
$\partial_{E^c}\Lambda_3^{ab} $&                      $-\partial_{E^c}\Lambda_2^{ab} $ &               $\partial_{E^c}\Lambda_4^{ab} $ &    $\partial_{E^c}\Lambda_2^{ab} $\\
$\partial_{E^c}\Lambda_3^{bb} $& 		      $-\partial_{E^c}\Lambda_2^{bb} $ & 	      $\partial_{E^c}\Lambda_4^{bb} $ &   $-\partial_{E^c}\Lambda_2^{bb} $\\
$\partial_{E^c}\Lambda_3^{cb} $& 		      $-\partial_{E^c}\Lambda_2^{cb} $ & 	      $-\partial_{E^c}\Lambda_4^{cb} $ &	   $\partial_{E^c}\Lambda_2^{cb} $\\
[0.1ex] \hline
$\partial_{E^c}\Lambda_4^{ab} $&                      $-\partial_{E^c}\Lambda_1^{ab} $ &               $\partial_{E^c}\Lambda_3^{ab} $ &    $\partial_{E^c}\Lambda_1^{ab} $\\
$\partial_{E^c}\Lambda_4^{bb} $& 		      $-\partial_{E^c}\Lambda_1^{bb} $ & 	      $\partial_{E^c}\Lambda_3^{bb} $ &   $-\partial_{E^c}\Lambda_1^{bb} $\\
$\partial_{E^c}\Lambda_4^{cb} $& 		      $-\partial_{E^c}\Lambda_1^{cb} $ & 	      $-\partial_{E^c}\Lambda_3^{cb} $ &	   $\partial_{E^c}\Lambda_1^{cb} $\\
[0.1ex] \hline
\end{tabular}
\caption{Transformation of tensor $\Lambda_{i}^{\mu b}$ and its derivatives with respect to the electric field under the generators of $Pnma$ (modulo a primitive translation). The superscript
indices label the coordinate while the subscript index labels the magnetic sublattice.}
\label{tab_Lambda}
\end{table}
The first two sets of equations in Eq.~(\ref{Eq_OL_DLambda}) together with Eq.~(\ref{Eq_OL_MuL}) and the sign of the components of the spins along $b$ at every magnetic sublattice shows that an
applied electric field along $\mathbf{a}$ ($\mathbf{b}$) induces a net orbital moment along $\mathbf{b}$ ($\mathbf{a}$).\\
In the same way, from the last equation in Eq.~(\ref{Eq_OL_DLambda}) one can derive that an applied electric field along $\mathbf{c}$ affects the orbital magnetic moments along $\mathbf{b}$ of
sublattice $1$ and $4$ in the opposite way to those of sublattices $2$ and $3$ giving rise to a zero net magnetization. %
\begin{figure}[htb]
 \includegraphics[scale=0.4]{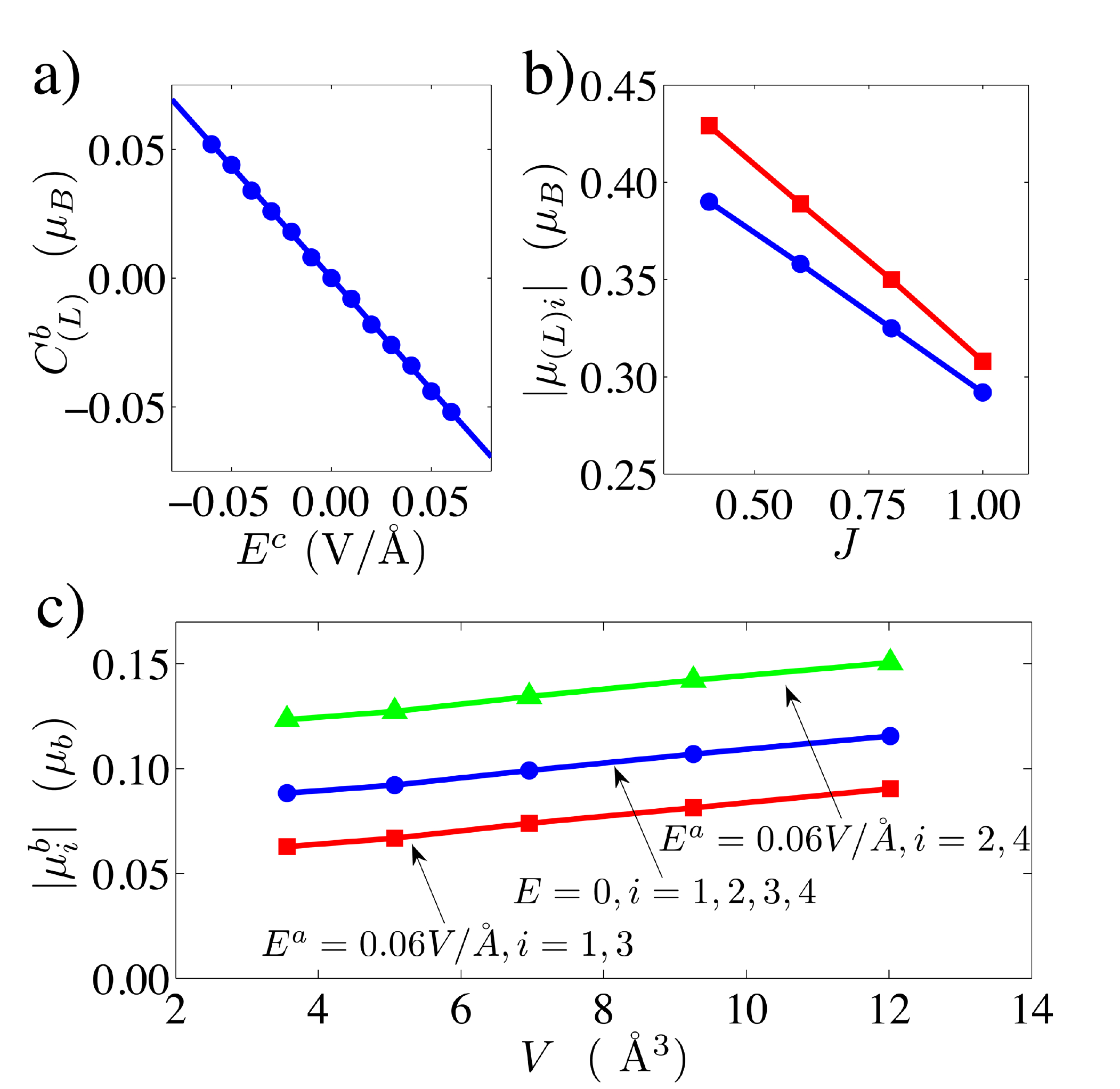}
\caption{Electric field dependence of $C^{b}_{(L)}$ (a) for $J=1$ eV (blue dots). (b) $J$ dependence of the size of orbital
moments at each magnetic sublattice for $U=4$ eV (blue dots) and $U=5$ eV (red dots). (c) Size of the orbital moment as a function of integration sphere volume for $E=0$ V/\AA (blue dots) and $E=0.06$
V/\AA  (green triangles and red squares). The lines are guides for the eyes. }
 \label{Fig_CLandJ}
\end{figure}
However, the staggered orbital magnetization $\mathbf{C}_{(L)} = \mbox{\boldmath$\mu$}_{(L)1} -\mbox{\boldmath$\mu$}_{(L)2}  -\mbox{\boldmath$\mu$}_{(L)3} +\mbox{\boldmath$\mu$}_{(L)4}$ along
$\mathbf{b}$ should by symmetry depend linearly on the applied electric field. 
To check this we perform the same calculations described in the text but for $\mathbf{E}\parallel\mathbf{c}$. The obtained electric field dependence of $C^{b}$ for $J=1$ eV is plotted in
Fig.~\ref{Fig_CLandJ}(a) and shows that such response is approximately double that of the orbital magnetization described in the text. 
In the same way we expect a linear dependence of $C^c_{(L)}$ on $\mathbf{E}\parallel \mathbf{b}$. However, such dependence is so weak that, as mentioned in the text, the changes in orbital moments are
above numerical resolution only at $E=0.06$ V/\AA .\\
We also note that the size of orbital moments strongly depends on the values of the on-site Coulomb interaction $U$ and on the effective on-site exchange parameter $J$. 
Figure~\ref{Fig_CLandJ}(b) shows the size of calculated orbital magnetic moment at $E=0$ as a function of $J$ for $U=4$ eV (blue dots) and $U=5$ eV (red dots). \\
{\it Convergence of the orbital moment.}
Finally, we analyze the convergence of the local orbital moments with respect to the radius of the  spheres in which they are evaluated. The modulation of this radius in VASP would require the
generation of new pseudo-potentials, which is not an easy task. Hence we decided to use an all-electron method based on linear augmented plane waves (LAPW), namely ELK~\cite{Elk}, which allows to
alter the radius of the projection spheres. First, we compared orbital moments using both methods and the same radius of the spheres finding that in the ground-state structure their values agree up to
10$^{-4}\mu_B$. Next, we varied the radius of the spheres in the ELK code. For this test, we used a GGA exchange correlation potential to avoid any additional influence of the Hubbard $U$ applied
within
the sphere. Figure~\ref{Fig_CLandJ}(c) shows the obtained value of orbital moments as a function of sphere volume at two Fe sublattice sites for the case of $E=0$ (blue squares) and $E^b=0.06 V/\AA$
(red squares and green triangles). 
While the orbital moment increases approximately linearly with sphere radius over the investigated range, the difference of the orbital moment is constant, so we can conclude that the response is well
converged even at the smallest sphere radii. 
It is worth mentioning that the orbital moment response obtained using GGA is smaller than that from with LDA+$U$. This finding may be attributed to a volume effect since the structure used was not
optimized for GGA.

{\it Additional technical details.}
All VASP density functional calculations were performed with an energy cut-off $E_{cut}=600$ eV. The ionic relaxation, the calculation of the force constant matrix and the calculations of the orbital
magnetism were done using a $2\times4\times4$ Monkhorst-Pack k-point mesh. 
The Born effective charges were obtained using both the finite electric field method (after relaxation using $4 \times 4\times 4$ gamma centered k-point mesh) and density functional perturbation
theory (with a $2\times 4\times 4$ Monkhorst-Pack k-point mesh). The results obtained with the two procedures had minimal differences.
The force constant matrix was obtained using finite displacements method obtained through the software Phonopy \cite{Phonopy}.
In the LAPW ELK calculation we used a $3\times5\times5$ k-point mesh and iterated the density until the change in potential was smaller than $10^{-6}$ eV.